\documentclass[twocolumn,showpacs,preprintnumbers,amsmath,amssymb,superscriptaddress, bibnotes]{revtex4}

\usepackage{color}
\usepackage{graphicx}
\usepackage{dcolumn}
\usepackage{bm}
\usepackage{mathrsfs}
\begin{document}

\title{Unconventional superconductivity and antiferromagnetic quantum critical behavior \\in the isovalent-doped  BaFe$_2$(As$_{1-x}$P$_x$)$_2$}

\author{Y.~Nakai}
\email{nakai@scphys.kyoto-u.ac.jp}
\author{T.~Iye}
\author{S.~Kitagawa}
\author{K.~Ishida}
\author{H.~Ikeda}
\affiliation{Department of Physics, Graduate School of Science, Kyoto University, Kyoto 606-8502, Japan,}
\affiliation{TRIP, JST, Sanban-cho Building, 5, Sanban-cho, Chiyoda, Tokyo 102-0075, Japan,}

\author{\\S.~Kasahara}
\author{H. Shishido}
\affiliation{Research Center for Low Temperature and Materials Sciences, Kyoto University, Kyoto 606-8502, Japan}
\author{T.~Shibauchi}
\author{Y.~Matsuda}
\affiliation{Department of Physics, Graduate School of Science, Kyoto University, Kyoto 606-8502, Japan,}
\author{T.~Terashima}
\affiliation{Research Center for Low Temperature and Materials Sciences, Kyoto University, Kyoto 606-8502, Japan}

\date{\today}

\begin{abstract}
{Spin dynamics evolution of BaFe$_2$(As$_{1-x}$P$_x$)$_2$ was probed as a function of P concentration via $^{31}$P NMR. Our NMR study reveals that two-dimensional antiferromagnetic (AF) fluctuations are notably enhanced with little change in static susceptibility on approaching the AF phase from the superconducting dome. Moreover, magnetically ordered temperature $\theta$ deduced from the relaxation rate vanishes at optimal doping. These results provide clear-cut evidence for a quantum-critical point (QCP), suggesting that the AF fluctuations associated with the QCP play a central role in the high-$T_c$ superconductivity.}

\end{abstract}

\maketitle
Conventional phonon-mediated superconductivity occurs in a normal metal that is well accounted for by Landau's Fermi-liquid (FL) theory. However, the standard FL theory appears to break down above $T_c$ in many ``exotic" superconductors characterized by unconventional pairing rather than a conventional uniform-sign $s$-wave pairing function, such as in the heavy fermion materials and cuprates. The origin of the anomalous metallic properties, often referred to as ``non-Fermi-liquid" (nFL) behavior, has triggered a growing interest in quantum-critical points (QCPs) that provide a route towards nFL behavior~\cite{LohneysenRMP}. The quantum-critical fluctuations induced by suppression of antiferromagnetic (AF) order~\cite{Sachdevtext} are suggested to mediate the Cooper pairing in exotic superconductors, in an analogous way to phonons in conventional superconductors~\cite{MathurNature}. 

Newly discovered iron-pnictide high-$T_c$ superconductivity also appears where antiferromagnetism is suppressed via chemical substitution or pressure~\cite{KamiharaFeAs}. The existence of a QCP in iron pnictides has been suggested~\cite{ChuPhysRevBBa(FeCo)2As2PhaseDiagram,NingContrastingSpinDynamics,CruzCeFe(AsP)OPRL,JDaiPNAS2009,KawasakiCaFe2As2QCP}. Spin-fluctuation-mediated superconductivity associated with the suppression of the antiferromagnetism is one likely scenario~\cite{MazinPRL2008}, but the identification of the mechanism is far from settled~\cite{KontaniOrbital,MazinPhysicaC2009Review}.
The difficulty in examining the superconducting (SC) mechanism could arise from complexity in the materials that can lead to ambiguous interpretations; e.g., non-universal SC gap functions and limited sample quality. It is thus essential to find a suitable model system to examine the mechanism of superconductivity. 

The isovalent-doped BaFe$_2$(As$_{1-x}$P$_x$)$_2$ can be used as such a model system. It has the highest $T_c$ (31 K) among iron-pnictide superconductors known to have line nodes in the SC gap~\cite{HashimotoPenetrationBaFe2(AsP)2,NakaiBaFe2(AsP)2,KimSpecificHeatBaFe2(AsP)2}. Clarifying the mechanism that produces its high-$T_c$ nodal gap is thus very important. Since isovalent P-doping is not expected to add carriers~\cite{KasaharaBaFe2(AsP)2}, BaFe$_2$(As$_{1-x}$P$_x$)$_2$ maintains the compensation condition, i.e. the volume of the hole Fermi surfaces (FSs) is equal to that of the electron FSs. Very clean single-crystals of BaFe$_2$(As$_{1-x}$P$_x$)$_2$ allow de Haas-van Alphen (dHvA) experiments that are the most precise technique to determine FSs, revealing the detailed electronic structure for comparison with band calculations~\cite{ShishidoBaFe2(AsP)2,AnalytisBaFe2(AsP)2}. The quasiparticle effective mass increases towards the maximum $T_c$, signaling the enhancement of electron-electron correlation. 
Such an increase in the quasiparticle mass as well as nFL behavior inferred from resistivity measurements~\cite{KasaharaBaFe2(AsP)2,JiangBaFe2(AsP)2} can be expected when the system is in proximity to a QCP. However, direct evidence for the existence of a QCP remains lacking in BaFe$_2$(As$_{1-x}$P$_x$)$_2$. 

Here, we report the normal-state spin dynamics in BaFe$_2$(As$_{1-x}$P$_x$)$_2$ for $0.2 \le x \le 0.64$ investigated by $^{31}$P NMR measurements. NMR is highly sensitive to low-energy spin fluctuations and can give information of the dynamical magnetic susceptibility. Samples of a mosaic of single crystals were prepared as described elsewhere~\cite{KasaharaBaFe2(AsP)2}. The dHvA experiments observed signals in single crystalline samples from the same batch as $x=$0.41, 0.56, and 0.64, indicating the excellent quality of our samples~\cite{ShishidoBaFe2(AsP)2}. The high quality of our samples is also suggested by the sharp SC transitions~\cite{KasaharaBaFe2(AsP)2}. $^{31}$P-NMR spectra ($^{31}\gamma/2\pi=17.237$ MHz/T) were obtained by sweeping frequency in a fixed magnetic field of 4.12 T. The Knight shift $K$ was determined with respect to the reference material H$_3$PO$_4$. The $^{31}$P nuclear spin-lattice relaxation rate $T_1^{-1}$ was determined by fitting the time dependence of spin-echo intensity after saturation of nuclear magnetization to a theoretical $I$ = 1/2 curve with a single component of $T_1$. 

Figure~1 (a) displays $^{31}$P NMR spectra obtained from the BaFe$_2$(As$_{1-x}$P$_x$)$_2$ crystals. Each $^{31}$P NMR spectrum consists of a single line, ruling out microscopic inhomogeneity caused by P substitution. 
\begin{figure}[tb]
\begin{center}
\includegraphics[width=7.5cm]{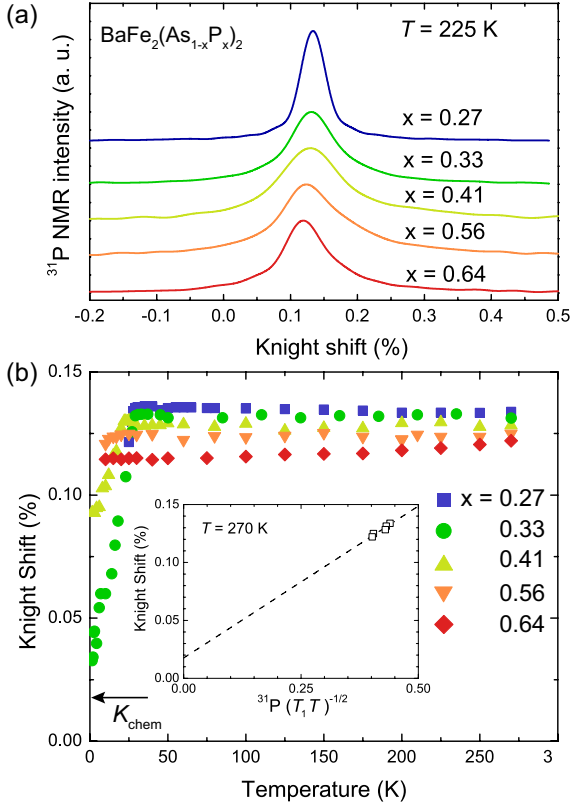}
\end{center}
\caption{(Color online) P substitution evolution of (a) $^{31}$P NMR spectra and (b) $^{31}$P Knight shift ($^{31}K$) determined at the spectral peak, obtained in a mosaic of the single crystals at 4.12 T. (b) The arrows indicate the chemical shift $K_{\rm chem}$ (see text). The abrupt decrease in $^{31}K$ at low temperatures is due to the onset of superconductivity. Inset:  $^{31}K$ vs $(T_1T)^{-1/2}$ at 270 K for different P concentrations $x$. 
}
\label{K}
\end{figure}
\begin{figure}[tb]
\begin{center}
\includegraphics[width=8cm]{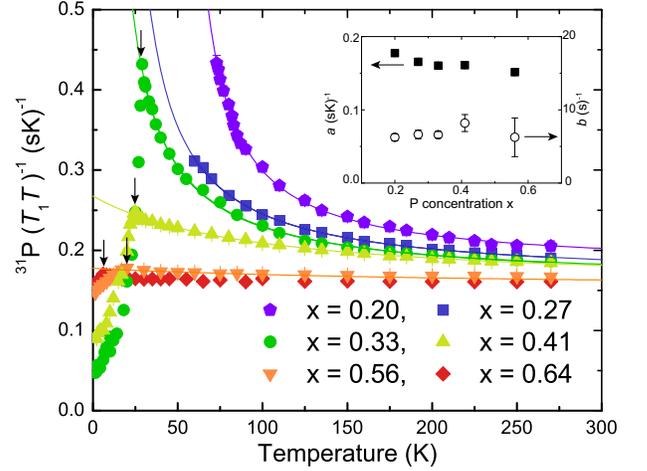}
\end{center}
\caption{(Color online) The $^{31}$P nuclear spin-lattice relaxation rate divided by temperature $(T_1T)^{-1}$ for BaFe$_2$(As$_{1-x}$P$_x$)$_2$ at 4.12 T. Solid lines represent fits to $(T_1T)^{-1} \!=\! a + b(T+\theta)^{-1}$ (see text). As AF fluctuations are suppressed as inferred from the suppression of $(T_1T)^{-1}$, $T_c$ (denoted by arrows) also decreases. Inset: Fitting parameters of $(T_1T)^{-1}$. The fitting parameters $a$ and $b$ are plotted against P concentration $x$. The $a$ and $b$ weakly depend on $x$, but $\theta$ shows a strong $x$ dependence (see Fig.~\ref{PhaseDiagram}). The small value of $\theta$ at $x$ = 0.33 ($\theta\sim0$) are insensitive to the fitting parameters of $a$ and $b$.}
\label{T1}
\end{figure}
$K$, which measures the effective field at the nucleus produced by electrons, is described as $K = K_{\rm spin} + K_{\rm chem}$; $K_{\rm spin}$ is the spin part of $K$ and is related to the uniform spin susceptibility $\chi({\bm q} = 0)$, which is proportional to the density of states at the Fermi energy $N(E_F)$. $K_{\rm chem}$ is the chemical shift, which is unrelated to $\chi({\bm q} = 0)$, and is estimated to be $\sim 0.018$ \% as follows. Since no obvious AF fluctuations were detected by NMR at high temperatures as seen in Fig.~2, it would be a good approximation to assume that $(T_1T)^{-1/2}$ is proportional to $N(E_F)$ at high temperatures, i.e. to assume that the usual Korringa relation holds at 270 K. Based on the plot of $(T_1T)^{-1/2}$ against $K$ at 270 K for different $x$ shown in the inset of Fig.~\ref{K}, we can estimate $K_{\rm chem}$ as the intercept. The obtained $K_{\rm chem}$ is $0.018\pm0.019$ \%, indicating that $K_{\rm spin}$ accounts for 86 \% of the observed Knight shift for $x=0.33$. Note that this $K_{\rm chem}$ would be a reasonable value, since the chemical shift for $^{31}$P in many diamagnetic insulators is of the order of some hundreds of ppm, which is comparable to this $K_{\rm chem}$~\cite{Carter}. By assuming $K_{\rm spin} \propto \chi({\bf q} = 0) = \mu_B^2N(E_F)$, the P-substitution dependence of $K_{\rm spin}$ at 270 K suggests that the change in $N(E_F)$ would be at most 10 \% for $x\le0.64$, which is quantitatively consistent with the result of our band calculation discussed below. 

Band-structure calculations by local-density-approximation (LDA) were performed for non-spin-polarized BaFe$_2$As$_2$ and BaFe$_2$P$_2$, using the WIEN2k package in the APW+local orbital basis~\cite{WIEN2k}. In addition, to obtain systematic changes of the electronic band structure for BaFe$_2$(As$_{1-x}$P$_x$)$_2$, we performed the LDA calculations for three virtual materials with linearly-interpolated $z=z_0(1-x)+z_1x$; (i) BaFe$_2$As$_2$ with fixed $(a,c)=(a_0,c_0)$, (ii) BaFe$_2$As$_2$ with linearly-interpolated $(a,c)=(a_0(1-x)+a_1x,c_0(1-x)+c_1x)$, (iii) BaFe$_2$P$_2$ with fixed $(a,c)=(a_1,c_1)$, where $a_{0,1}$, $c_{0,1}$, and $z_{0,1}$ are the experimental values for the crystallographic parameters of BaFe$_2$As$_2$ and BaFe$_2$P$_2$, respectively~\cite{KasaharaBaFe2(AsP)2,RotterPRB2008}. $N(E_F)$ barely changes for $x<0.5$, and then decreases for $x>0.5$ (see supplementary information). Such behavior is consistent with our Knight shift results. 

As shown in Fig~\ref{K}~(b), $K$ is almost temperature independent for $x\le0.56$, and that the absolute value of $K$ in the normal state decreases only slightly upon P substitution. These data indicate that P substitution does not produce significant changes in $\chi({\bf q} =0)$ and $N(E_F)$. 
This is in stark contrast to carrier-doped iron-pnictide superconductors; in electron-doped Ba(Fe$_{1-y}$Co$_y$)$_2$As$_2$, the Knight shift data indicate that $N(E_F)$ of non-SC $y$ = 0.26 is approximately 50 \% that of $y$ = 0.08 with the maximum $T_c$ of 26 K~\cite{NingContrastingSpinDynamics}. Such drastic effects on $N(E_F)$ via electron-doping is expected from the characteristic band structure~\cite{SinghPRL2008}; the calculated $N(E_F)$ rapidly changes near $E_F$ with a negative gradient, resulting in a rapid decrease of $N(E_F)$ with electron-doping. Large changes in FSs via charge-carrier doping thus necessarily involve dramatic modification in the $N(E_F)$ and Fermi-surface nesting resulting in changes in spin fluctuations. In addition to possible changes in $T_c$ due to the modification of spin excitation spectrum~\cite{NingContrastingSpinDynamics}, drastic changes in $N(E_F)$ can also affect severely $T_c$~\cite{KuchinskiiDOSvsTc}, and the suggested giant magnetoelastic coupling may lead to further suppressions of $T_c$~\cite{YildirimPhysicaC2009}. Therefore, the decrease in $N(E_F)$ as well as the suppression of spin fluctuations should be taken into account for the interpretation of possible changes in $T_c$ for electron-doped Ba(Fe$_{1-y}$Co$_y$)$_2$As$_2$. In contrast, the nearly unperturbed $K_{\rm spin}$ by isovalent P-doping demonstrates that BaFe$_2$(As$_{1-x}$P$_x$)$_2$ is an ideal model system to test the relevance of spin fluctuations to superconductivity. 

Significant low-energy AF fluctuations are probed near the maximum $T_c$ via $T_1^{-1}$. $(T_1T)^{-1}$ is described by the wave-vector average of the imaginary part of the dynamical susceptibility $\chi^{\prime\prime}({\bf q},\omega_0)$, i.e., $(T_1T)^{-1}\propto\sum_{\bf q}|A({\bf q})|^2\chi^{\prime\prime}({\bf q},\omega_0)/\omega_0$ where $A({\bf q})$ represents the hyperfine coupling between $^{31}$P nuclear spins and the surrounding electrons, and $\omega_0$ is NMR frequency. The Korringa law $T_1TK^2 = const.$ generally holds in a FL state, but is inapplicable to BaFe$_2$(As$_{1-x}$P$_x$)$_2$ near the AF phase since Curie-Weiss (CW) behavior is observed in $(T_1T)^{-1}$ (see Fig.~\ref{T1}). For $x=0.33$, $(T_1T)^{-1}$ increases significantly down to $T_c$ whereas the Knight shift is constant. These $(T_1T)^{-1}$ and $K$ data demonstrate convincingly that AF fluctuations with finite {\bf q} continue to grow down to $T_c$ at optimal doping~\cite{NakaiBaFe2(AsP)2}.

\begin{figure}[tb]
\begin{center}
\includegraphics[width=7.5cm]{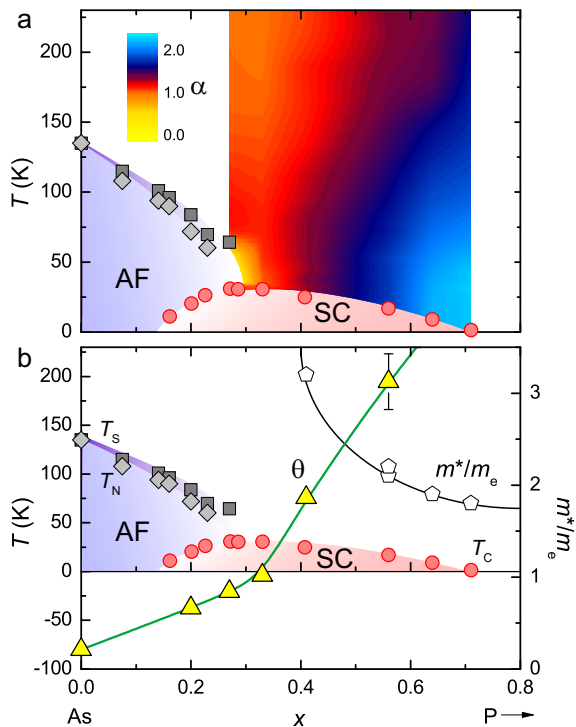}
\end{center}
\caption{(Color online) Temperature - P concentration phase diagram of BaFe$_2$(As$_{1-x}$P$_x$)$_2$. 
$T_c$, $T_s$, and $T_N$ denote the SC, orthorhombic-tetragonal, and AF transition temperatures, respectively. 
(a) Colors represent the exponent, $\alpha = d(\ln{\Delta\rho_{ab}})/d(\ln{T})$, at zero magnetic field for the resistivity curves in Ref.~24, 
where $\Delta\rho_{ab} = \rho_{ab}-\rho_{ab}(T \to 0$ K). 
(b) The triangles represent $\theta$ obtained by fitting $(T_1T)^{-1}$ data to the equation, $(T_1T)^{-1} = a + b(T + \theta)^{-1}$ (error bars are within the symbol size). 
$T$-linear resistivity, suggesting quantum-critical behavior, is observed in proximity to a magnetic instability signaled by $\theta\simeq 0$ at $x=0.33$. 
These quantum-critical fluctuations are found to correlate with the enhancement of the effective mass as denoted by the pentagons (taken from Ref.~17). 
}
\label{PhaseDiagram}
\end{figure}

The AF fluctuations in BaFe$_2$(As$_{1-x}$P$_x$)$_2$ are enhanced significantly as the P concentration is reduced towards the maximum $T_c$ ($x\simeq0.33$), as evidenced by the rapid increase in $(T_1T)^{-1}$ from conventional FL behavior at $x$ = 0.64 (where $(T_1T)^{-1}$ and $K$ are almost constant). The crossover from FL to CW behavior in $(T_1T)^{-1}$ correlates perfectly with the change in the resistivity results~\cite{KasaharaBaFe2(AsP)2}; As the system evolves from a Fermi liquid ($x=0.71$) towards the maximum $T_c$ ($x=0.33$) near the AF phase, the temperature dependence of the resistivity changes from $T^2$ to $T$-linear, one hallmark of nFL behavior. 
The exponent of the temperature dependence of the resistivity is shown as a contour plot in Fig.~\ref{PhaseDiagram} (a). Specifically, the CW behavior of $(T_1T)^{-1}$ and the $T$-linear resistivity at $x$ = 0.33 can be explained by the existence of two-dimensional (2D) AF spin fluctuations in the theory of nearly AF metals~\cite{MoriyaTakahashiUedaJPSJ1990}. Indeed, the evolution of the AF spin excitations measured by $(T_1T)^{-1}$ upon P substitution can be fit consistently by the equation expected from the same theory~\cite{MoriyaTakahashiUedaJPSJ1990}, $(T_1T)^{-1}=a + b(T+\theta)^{-1}$ (see solid lines in Fig.~\ref{T1}). Such 2D AF fluctuations were experimentally observed in the parent BaFe$_2$As$_2$ via neutron scattering experiments~\cite{MatanPRB2009}. According to band calculations~\cite{HashimotoPenetrationBaFe2(AsP)2}, substantial 2D AF fluctuations can be generated by the inter-band spin excitations between the multiple FSs predominantly derived from Fe $d$ electrons.

Our central finding in BaFe$_2$(As$_{1-x}$P$_x$)$_2$ is that the 2D AF fluctuations of a quantum-critical nature have a clear correlation with the enhancement of quasiparticle effective mass and $T_c$ as summarized in Fig.~\ref{PhaseDiagram}. We found that the Weiss temperature $\theta$ obtained from the fitting increases with P substitution and becomes almost zero near $x$ = 0.33 where the maximum $T_c$ is achieved. $\theta=0$ K implies that the dynamical susceptibility probed by $(T_1T)^{-1}$ measurement diverges at absolute zero, or that the magnetic correlation length continues to increase down to $T = 0$ K. Our result thus strongly suggests the presence of an AF QCP near the maximum $T_c$ in proximity to the AF phase boundary. As the P concentration is varied towards optimal doping ($x\sim0.33$) from $x = 0.64$ where the FL state is observed, the magnetic fluctuations become dramatically enhanced as $\theta$ decreases. Importantly, the quasiparticle mass and $T_c$ increase as $\theta$ approaches 0 K. This strongly suggests that the AF quantum-critical fluctuations lead to strong mass renormalization and unconventional superconductivity. Our systematic NMR measurements, which are compared with transport measurements, thus provide the first evidence that the quasiparticle mass enhancement is strongly coupled to the AF quantum-critical fluctuations in iron-pnictide superconductors as previously observed in heavy fermion systems~\cite{ShishidoJPSJCeRhIn5}. Furthermore, since $N(E_F)$ generally correlates with $T_c$ in conventional BCS superconductors, the enhancement of $T_c$ with approaching the QCP from the overdoped side cannot be accounted for by an nearly unperturbed $N(E_F)$, demonstrating clearly that superconductivity in BaFe$_2$(As$_{1-x}$P$_x$)$_2$ is tuned predominantly by the AF fluctuations. Therefore, we conclude that the 2D AF quantum-critical fluctuations are likely to play a central role in the occurrence of unconventional superconductivity in BaFe$_2$(As$_{1-x}$P$_x$)$_2$. 

Previous NMR measurements also indicated that there is a strong connection between low-energy spin fluctuations and superconductivity in 122~\cite{NakaiBaFe2(AsP)2}, 111~\cite{LMa_NaFeAsP} and 11 iron-based superconductors~\cite{ImaiFeSePRL}. These suggest that the relevance of spin fluctuations to the high-$T_c$ superconductivity is universal despite the non-universal gap structure. The strong connection would be naturally understood if one considers the superconductivity observed near magnetic phase and the well-nested hole and electron Fermi surfaces. By contrast, we reported that the connection is relatively weaker in LaFeAs(O$_{1-x}$F$_x$)~\cite{NakaiJPSJ2008}. Such a weak link may originate from much lower $T_c$ of 26 K in LaFeAs(O$_{1-x}$F$_x$) than that of $R$FeAs(O$_{1-x}$F$_x$) ($R$ = Ce, Pr, Sm, etc.). However, it remains unresolved whether spin fluctuations are also important for high-$T_c$ superconductivity in the 1111 system with $T_c$ exceeding 50 K, since magnetic rare-earth atoms $R$ hinder quasiparticle excitations responsible for superconductivity~\cite{YamashitaNMRReFeAsO1-dPhysica}. Therefore, examinations of the connection between magnetic fluctuations and superconductivity of high-$T_c$ 1111 iron pnictides would be crucial for establishing general view of iron pnictide superconductivity.

We thank K.~Kitagawa, Y.~Ihara, D.~C.~Peets and Y.~Maeno for experimental support and discussion. We also grateful to K.~Kuroki, H.~Kontani, Q.~Si and S.~Fujimoto for theoretical discussions. This work was supported by Grants-in-Aid for Scientific Research on Innovative Areas ``Heavy Electrons" (No. 20102006) from MEXT, for the GCOE Program ``The Next Generation of Physics, Spun from Universality and Emergence" from MEXT, and for Scientific Research from JSPS.

\end{document}